\documentstyle[prl,aps,multicol]{revtex}
 \begin{document}
 \draft
\title{Quantizing magnetic field and quark-hadron phase transition in 
a neutron star}
\author{Debades Bandyopadhyay,$^1$ Somenath Chakrabarty,$^2$ 
and Subrata Pal$^1$ }
\address{$^1$Saha Institute of Nuclear Physics, 1/AF Bidhannagar, 
Calcutta 700 064, India}
\address{$^2$Department of Physics, University of Kalyani, 
Kalyani 741235, India and IUCAA, P.B. 4, Ganeshkhind, Pune 411007, India}

\maketitle

\begin{abstract}
We investigate the influence of a strong magnetic field on various properties 
of neutron stars with quark-hadron phase transition. The one-gluon exchange
contribution in a magnetic field is calculated in a relativistic
Dirac-Hartree-Fock approach. In a magnetic field of $5\times 10^{18}$G in 
the center of the star, the overall equation of state is softer in 
comparison to the field-free case resulting in the reduction of maximum 
mass of the neutron star.
\end{abstract}


\begin{multicols}{2}

The matter density in the core of a neutron star could exceed up to a few
times the nuclear matter saturation density. At such high density, it is 
expected that the quark degrees of freedom would be manifested. In fact,
quark matter composed of comparable proportions of up, down and strange 
quarks has been conjectured \cite{Wit,Far} to be the true ground 
state of QCD at finite baryon density. Therefore, at such high baryon 
density, a transition from nuclear matter to a stable quark matter
is a possibility. Several authors \cite{Col,Gle,Pra,Mul} have studied 
the effect of this phase transition on neutron star properties. 

The presence of strong magnetic fields in neutron stars might have 
interesting astrophysical implications. Large magnetic field 
$B_m \sim 10^{14}$G has been estimated at the surface of neutron stars
\cite{Wol}. On the other hand, in the core, the field may have been 
amplified considerably due to flux conservation from the original weak 
field of the progenitor during its core collapse. In fact, field as large 
as $\sim 10^{18}$G in the core is predicted \cite{Lai} using scalar virial 
theorem which is based on Newtonian gravity. At such high matter density, 
the effect of general relativity is significant and this gives rise to a 
very strong gravitational force \cite{Sha} on the star. Consequently, the 
value of $B_m$ is expected to be further increased above $10^{18}$G. 
Because of highly conducting core, such a high field is frozen in \cite{Mus} 
and may not manifest at the surface. The energy of a charged particle 
changes significantly in the quantum limit if the magnetic field is comparable 
to or above a critical value $B_m^{(c)}$ \cite{Can}, and the quantum effects 
are most pronounced when the particle moves in the lowest Landau level. The 
interaction of charged particles with strongly quantizing fields has been 
shown to modify the gross properties of matter on the surface \cite{Lai,Yak} 
as well as in the core of neutron stars \cite{Vsh,Cha}.

Theoretical studies of Fock (exchange) term, relevant to the star 
surface/crust, in intense magnetic field have been carried out using 
the simple Thomas-Fermi-Dirac model \cite{Fus92}.
In this Letter, we investigate the composition and structure of neutron
stars with quark-hadron phase transition under the 
influence of strong magnetic fields in the Dirac-Hartree-Fock (DHF) 
approach within a mean-field approximation. This method is rather general, 
so it should be of correspondingly broad interest.

We describe the calculation of the Fock term in presence of a magnetic field 
in a general formalism within the $\sigma$-$\omega$ model \cite{Cha,Ser}. 
In a uniform magnetic field $B_m$ along z-axis, the Lagrangian is given by
\begin{eqnarray}
{\cal L} &=& {\bar \psi} \left[ i\gamma_{\mu}D^{\mu} - m - g_{\sigma}\sigma 
- g_{\omega}\gamma_{\mu}\omega^{\mu} \right] \psi + {1\over 2} 
(\partial^{\mu}\sigma)^2 \nonumber \\
&-& {1\over 2} m^2_{\sigma}\sigma^2 
- {1\over 4} \left( \partial_{\mu}\omega_{\nu} 
- \partial_{\nu}\omega_{\mu} \right)^2 + 
{1\over 2} m_{\omega}^2 (\omega_{\mu})^2 ,
\end{eqnarray}
in the usual notation \cite{Cha}. The general solution for protons is 
$ \psi({\bf r}) \propto \exp (-i\epsilon^{HF} t + ip_y y + ip_z z) 
f_{p_y,p_z}(x)$, where the 4-component spinor at zero temperature
$f_{p_y,p_z}(x)$ is of the same form as in Ref. \cite{Cha}, but with the 
single-particle Hartree energy $\epsilon^H$ replaced by the corresponding 
Hartree-Fock energy $\epsilon^{HF}$.
The form of the spinor in a magnetic field (see Ref. \cite{Cha})
restricts the evaluation of the Fock term to strong fields such
that only the zeroth Landau level, $\nu=0$, is populated. The position 
dependent part can then be decoupled so that it reduces to the form
\begin{equation}
f^{\nu=0}_{p_y,p_z}(x) = N_{\nu=0} \left( \matrix{ \epsilon^{HF}_{\nu=0} 
+ p_{vz}\cr  0\cr  - m^*\cr  0\cr} \right) I_{{\nu=0};p_y}(x)  , 
\end{equation}
where $N_{\nu=0} = 1/\sqrt{2\epsilon^{HF}_{\nu=0}(\epsilon^{HF}_{\nu=0} + p_{vz})}$ 
and $\epsilon^{HF}_{\nu=0} = \epsilon^{HF}_{p_z} - U^H_0 - U^F_0(p_z)
= \sqrt{p_{vz}^2 + {m^*}^2}$. 
The DHF equation of protons for $\nu=0$ can then easily
be written as
\begin{equation}
\left[ \alpha_z p_z + \beta \left( m + U^H + U^F \right) \right] u(p_z) 
= \epsilon^{HF}_{p_z} u(p_z) ,
\end{equation}
where the effective mass $m^* = m + U^H_S + U^F_S(p_z)$, 
and $u(p_z)$ is the momentum dependent part of the spinor. 
For $B_m \neq 0$, the Hartree 
contribution $U^H$ is given in Ref. \cite{Cha}. The Fock term $U^F$
in general is given by \cite{Leu} $\beta U^F(p_z) = \beta U^F_S(p_z) + 
U^F_0(p_z) + \mbox{\boldmath $\alpha$} \cdot {\hat p} U^F_V(p_z)$,
where $\hat p$ is the unit vector along $B_m$. For $B_m \neq 0$ the 
different terms are given by 
$$ U^F_S(p_z) = {1\over 16\pi^2} \int^{+p_F}_{-p_F} dq_z \; 
{m^*\over \sqrt{q^2_{vz} + m^{* 2}}} \left( {\cal J}_{\sigma} 
- 4{\cal J}_{\omega} \right), $$
$$ U^F_0(p_z) = {1\over 16\pi^2} \int^{+p_F}_{-p_F} dq_z 
\left( {\cal J}_{\sigma} + 2{\cal J}_{\omega} \right), $$
$$ U^F_V(p_z) = - {1\over 16\pi^2} \int^{+p_F}_{-p_F} dq_z \; 
{q_{vz}\over \sqrt{q^2_{vz} + m^{* 2}}} \left( {\cal J}_{\sigma} 
+ 2{\cal J}_{\omega} \right). $$
Here $q_{vz} = q_z\left(1 + U^F_V(p_z)/q_z\right)$ and
${\cal J}_a = g_a^2 \exp\left( l_a^2/2qB_m\right) \int^{\pi/2}_0 
d\theta \sec\theta \; {\rm erf} \left( l_a \sec\theta/ \sqrt{2qB_m} \right)$
with $l_a^2 = \left( p_z - q_z \right)^2 - \left( \epsilon^{HF}_{p_z} - 
\epsilon^{HF}_{q_z} \right)^2 + m_a^2$, and ${\rm erf}(x)$ denotes the
error function; $g_a$s correspond to meson coupling constants with 
$a \equiv (\sigma,\omega)$. The exchange contribution for $B_m = 0$ is 
given in Ref. \cite{Leu,Jam}. It is straightforward to extend this 
formalism to calculate the exchange interaction for electrons in the
outer crust of a strongly magnetized neutron star \cite{Fus92,Ban} 
and also for a magnetized pair Fermi gas \cite{Dai} at finite temperature 
and baryon density in exotic stellar objects and cosmology. To 
explore the quark-hadron phase transition, we demonstrate here the extension 
of the DHF approach to the calculation of the one-gluon exchange term 
in the quark phase. 

The pure quark phase consisting of $u$, $d$ and $s$ quarks interacting
through one-gluon exchange in local charge neutral and $\beta$-equilibrium
conditions is described by the bag model \cite{Far}. The DHF 
equations of motion for quarks in strong magnetic field can readily 
be obtained from Eq. (3) by dropping the $\sigma$-meson term and replacing
the $\omega$-meson coupling by the quark-gluon coupling, i.e. 
$g_{\omega} \to (g/2) \lambda_{ab}^{\gamma}$, where $a(b)$ corresponds to the 
color charge of the outgoing(incoming) quark and $\lambda$s are SU(3) generators. 
The Hartree term ($U^H$) vanishes because the color symmetric combination 
($\sim {\rm tr} \lambda$) does not couple to the gluons.
The interaction energy density is then due to exchange term only, and this 
to order $g^2$ for each flavor with $B_m \neq 0$ is 
$$ {\cal E}^{f;\nu=0}_I = {q_f B_m\over 8\pi^2} \int^{+p_F^f}_{-p_F^f} dp_z
\left[ U^F_0 + {p_{vz}\over \sqrt{p^2_{vz} + m_f^{* 2}}} U^F_V \right], $$
where $q_f$, $m^*_f$ and $p_F^f$ are the charge, effective mass and 
Fermi momentum of quark of $f$th flavor.
The QCD coupling constant is defined by $\alpha_c = g^2/4\pi$. The general 
expression (for all $\nu$) for the total kinetic energy of the quark 
phase in a magnetic field is
\begin{eqnarray}
{\cal E}^{\nu}_K &=& \sum_{f=u,d,s} {d_fq_fB_m\over 4\pi^2}
\sum^{\nu_{\rm max}^{(f)}}_{\nu=0} g_{\nu} 
\Phi\left(\mu_f^*,m^*_{f,\nu}\right) \nonumber\\
&+& {eB_m\over 4\pi^2} \sum^{\nu_{\rm max}^{(e)}}_{\nu=0} g_{\nu}
\Phi\left(\mu_e,m_{e,\nu}\right),
\end{eqnarray}
where $\Phi(x,y) = x{\cal O}_{i,\nu}^{1/2} + y^2 \; {\rm ln} \left\{ 
\left( x + {\cal O}_{i,\nu}^{1/2}\right) \Big/y \right\}$  with $i\equiv f$ or $e$.
The notation in Eq. (4) is same as that in Ref. \cite{Cha}, but the first term 
corresponds to those for quarks with $d_f=3$ and the second term is
that for electrons. The total energy density of the pure quark phase is then 
${\cal E}^q = {\cal E}^{\nu}_K + \sum_f {\cal E}_I^{f;\nu=0} + {\cal E}_m + B$, 
where $B$ is the bag constant and the magnetic field energy density is
${\cal E}_m = B_m^2(n_b/n_0)/(8\pi)$. Since the higher order contributions 
of the density dependent field $B_m(n_b/n_0)$ (given by Eq. (5)) to number 
densities and chemical potentials of different species are found to be 
negligible, the magnetic energy density and magnetic pressure have been 
treated perturbatively. The pressure follows from the relation
$P^q = \sum_f \mu_f n_f + \mu_e n_e - {\cal E}^q$, where $\mu_f$ denotes  
the quark chemical potential and 
$n_f = \left( d_fq_fB_m/2\pi^2 \right) \sum_{\nu=0}^{\nu_{\rm max}^{(f)}} 
g_{\nu} \left( \mu_f^{* 2} - m_{f,\nu}^{* 2} \right)^{1/2}$ is the quark 
density.  The electron density is $n_e = \left( eB_m/2\pi^2 \right) 
\sum_{\nu=0}^{\nu_{\rm max}^{(e)}} g_{\nu} \left( \mu_e^2 
- m_{e,\nu}^2 \right)^{1/2}$, and $\mu_e$ its chemical potential.
The charge neutrality condition, 
$Q^q = \sum_f q_f n_f - n_e = 0$, and the $\beta$-equilibrium conditions,
$\mu_d = \mu_u + \mu_e = \mu_s$, can be solved self-consistently together with
the effective masses at a fixed baryon number density $n_b^q = (n_u+n_d+n_s)/3$ 
to obtain the equation of state (EOS) for the deconfined phase.
For the ease of numerical computation we, however, add here the one-gluon 
exchange term perturbatively to energy density and pressure.

To describe pure hadronic matter consisting of neutrons ($n$), protons ($p$) 
and electrons ($e$), we employ the linear $\sigma$-$\omega$-$\rho$ model 
of Ref. \cite{Zim} in the relativistic Hartree approach. Neglecting the Fock term 
in hadron phase is quite justified as the Hartree energy grows like $p_F^6$ 
while the exchange energy behaves as $p_F^4$ \cite{Chi}, so the Fock correction
to the EOS is expected to be small \cite{Jam,Chi}. The EOS for this phase is 
obtained by solving self-consistently the effective mass in conjunction with 
the charge neutrality and $\beta$-equilibrium conditions, $Q^h = n_p - n_e = 0$ 
and $\mu_n = \mu_p + \mu_e$ at a fixed baryon number density $n_b^h$. Here $n_i$ 
and $\mu_i$ denote the number density and chemical potential; the subscript $i$ 
refers to $n$, $p$ and $e$. The energy density 
${\cal E}^h= {\cal E}_0 + {\cal E}_m$ (${\cal E}_0$ is the contribution to the 
energy density due to nucleons and electrons \cite{Cha}) 
and pressure $P^h$ in this phase are related by 
$P^h = \sum_i n_i\mu_i - {\cal E}^h$.

The mixed phase of hadrons and quarks comprising of two conserved charges,
baryon number and electric charge is described following Glendenning \cite{Gle}.
The conditions of global charge neutrality and baryon number conservation 
are imposed through the relations $\chi Q^h + (1-\chi)Q^q = 0$ and 
$n_b = \chi n_b^h + (1-\chi)n_b^q$, where $\chi$ represents the fractional
volume occupied by the hadron phase. Furthermore, the mixed phase 
satisfies the Gibbs' phase rules: $\mu_p = 2\mu_u + \mu_d$ and $P^h = P^q$.
The total energy density is ${\cal E} = \chi{\cal E}^h + (1-\chi){\cal E}^q$. 

In the present calculation the values of the dimensionless coupling constants 
for $\sigma$, $\omega$ and $\rho$ mesons determined by reproducing the nuclear 
matter properties at a saturation density of $n_0 = 0.16$ ${\rm fm}^{-3}$ are 
adopted from Ref. 

 \end{multicols}
\begin{table}

\caption{ The phase boundaries, $u_1$ and $u_2$, and central densities $u_c$ 
of stars with maximum masses $M_{\rm max}/M_{\odot}$ with 
and without magnetic fields that undergo a quark-hadron phase 
transition with interacting quark phase. The corresponding quantities with 
non-interacting quark phase are shown in parentheses. The variation of 
magnetic field with density $n_b$ is given by Eq. (5) with $\beta = 0.01$, 
$\gamma =3$ for $B_0=5\times 10^{18}$G and $B_0=10^{19}$G. Calculations are 
performed for the hadronic Lagrangian of Zimanyi and Moszkowski (ZM) [21] and 
Serot and Walecka (SW) [16]. The bag constant is $B=250$ 
MeV ${\rm fm}^{-3}$ and the nuclear matter saturation density $n_0$ is 
0.16(0.1484) ${\rm fm}^{-3}$ for ZM(SW) model. }

\begin{tabular}{cccccc} 

Hadronic Model& $B_m$ (Gauss)& $u_1=n_1/n_0$& $u_2=n_2/n_0$& 
$u_c=n_c/n_0$& $M_{\rm max}/M_{\odot}$ \\ \hline
ZM& 0& 4.107(3.329)& 25.941(17.681)& 7.152(7.754)& 1.707(1.610) \\
\hfil& $10^{14}-5\times 10^{18}$& 4.108(3.326)& 25.754(17.530)& 
6.902(7.104)& 1.549(1.466) \\ 
\hfil& $10^{14}-10^{19}$& 4.130(3.312)& 25.383(17.085)& 5.214(5.016)& 
1.331(1.268) \\ \hline
SW& 0& 2.158(1.953)& 9.523(7.837)& 4.259(4.012)& 2.594(2.279) \\
\hfil& $10^{14}-5\times 10^{18}$& 2.159(1.945)& 9.374(7.769)& 
3.252(3.554)& 2.342(2.017) \\ 
\end{tabular}
\end{table}
\begin{multicols}{2}

\noindent \cite{Gle}. The current masses of $u$ and $d$ quarks 
are taken as $m_u = m_d = 5$ MeV and $m_s = 150$ MeV, and the QCD coupling 
constant is $\alpha_c = 0.2$. We consider the bag constant 
$B=250$ MeV ${\rm fm}^{-3}$ which corresponds to the lower limit dictated 
by the requirement that, at low density, hadronic matter is the preferred phase.
The variation of the magnetic field $B_m$ with density $n_b$
from the center to the surface of a star is parametrized by the form
\begin{equation} 
B_m(n_b/n_0) = B_m^{\rm surf} + B_0 \left[
1 - \exp\left\{ -\beta (n_b/n_0)^{\gamma} \right\} \right] ,
\end{equation} 
where the parameters are chosen to be $\beta = 0.01$ and $\gamma = 3$.
The maximum field prevailing at the center is taken as
$B_0=5\times 10^{18}$G and the surface field is 
$B_m^{\rm surf} \simeq 10^{14}$G. The number of Landau levels populated for 
a given species is determined by the field $B_m$ and baryon density \cite{Cha}. 
Over the entire density range from the surface to the core, we find that the 
nucleonic energy density (pressure) dominates substantially over the magnetic 
field energy density (pressure). Hence matter can sustain such a high field 
without injecting any instability in the corresponding EOS.  

With this parameter set, we show in Table I, the mixed phase boundaries of 
neutron star for $B_m=0$ and for $B_m \neq 0$ with 
interacting quark phase (IQP). To examine the measure of the importance of 
the exchange contribution, the corresponding results for matter with 
non-interacting quark phase (NQP) are given in parentheses. 
The onset of transition is at density $n_1=u_1n_0$ and the pure quark 
phase begins at density $n_2=u_2n_0$. For matter with NQP 
and $B_m=0$, the boundaries are at $u_1=3.329$ and 
$u_2=17.681$. With the 
inclusion of interaction, the EOS for the quark phase becomes softer and 
transition to a mixed phase is delayed to $u_1=4.107$. At high density, 
as expected, the EOS for the quark phase is more softer resulting in a 
larger shift in $u_2$ to 25.941 compared to that in $u_1$. As a consequence, 
the extent of the mixed phase is increased. With further inclusion of 
magnetic field $B_0= 5\times 10^{18}$G, the situation is reversed. The EOS for the 
hadronic sector for $B_m \neq 0$ is found to be softer at low density and 
it turns out to be stiffer at high density compared to the field-free case 
\cite{Cha}. On the other hand, the EOS for the IQP 
with $B_m \neq 0$ is found to be considerably stiffer over the entire 
density range compared to those for $B_m=0$, with or without the exchange 
term. Therefore, for $B_0 = 5\times 10^{18}$G, the onset of mixed phase 
occurs at 4.108. The stiff EOS for both hadronic
and interacting quark sectors at $B_0 =5\times 10^{18}$G reduce the
upper boundary to a density of $u_2=25.754$ and thereby also the extent 
of the mixed phase. Since the EOS of NQP with $B_0 = 5\times 10^{18}$G 
is most stiff of all the cases considered above, the boundaries
of the mixed phase are maximally shifted to the lowest baryon densities
at $u_1=3.326$ and $u_2=17.530$, and thus has the smallest mixed phase extent.

For the field-free case, with the appearance of quarks,
the neutron and electron abundances are found to decrease since 
quark matter furnishes both baryon number and negative charge. With 
$B_0=5\times 10^{18}$G,
apart from reducing the mixed phase extent, the magnetic field enhances
the electron fraction in the hadronic sector \cite{Cha}. Because of charge
neutrality condition $n_p=n_e$ in the hadronic phase the proton fraction 
is also increased. The enhanced electron abundance persists even in the 
mixed phase. In the mixed phase, with $B_0 = 5\times 10^{18}$G, the 
$u$-quark abundance is found to be enhanced, while those of $d$ and 
$s$ quarks remains practically unaltered.

The maximum masses of the stars $M_{\rm max}$, obtained by solving 
the TOV equation \cite{Sha} are given in Table I for different cases studied.
With only nucleons and electrons, without any quark, the maximum masses 
of the stars are $1.778M_{\odot}$ and $1.643M_{\odot}$ for $B_m=0$ and 
$B_0=5\times 10^{18}$G. This is a manifestation of the softening of the EOS 
in magnetic field for these stars \cite{Cha}. However, the introduction of 
quarks, which softens the overall EOS, causes the maximum mass to be smaller.
For $B_m=0$, the inclusion of one-gluon exchange in the quark sector causes a 
delayed appearance of the pure quark phase, and this results in a stiffer 
overall EOS with a larger maximum mass of $1.707M_{\odot}$ in contrast to 
matter with NQP. Due to further softening of the EOS by $B_0=5\times 10^{18}$G 
for matter with IQP, the maximum mass is reduced to a value of $1.549M_{\odot}$. 
The smallest $M_{\rm max} = 1.466M_{\odot}$ is obtained for the softest EOS with 
this value of $B_0$ for matter with NQP.
The central densities $u_c=n_c/n_0$ of the maximum mass stars for all cases
(see Table I) are less than $u_2$ and fall within the mixed 
quark-hadron phase. Therefore the presence of a pure quark phase is precluded. 

We have found that when the QCD coupling constant $\alpha_c > 0.25$, 
``strange matter" becomes negatively charged and requires positrons for 
overall charge neutrality and therefore would have catastrophic 
consequences \cite{Far}. 

Calculations are repeated with a higher value of $B_0 =10^{19}$G
(see Table I) \cite{Ins}. Here the maximum masses of the stars for both 
IQP and NQP are found to be smaller than the observational lower limit 
of $1.44M_{\odot}$ imposed by the larger mass of the binary pulsar 
PSR 1913+16 \cite{Wei}. 
Therefore, the quark-hadron phase transition with the present hadronic 
EOS \cite{Zim} sets a limit on the maximum value of the field that can be 
used for a stable system. To estimate the uncertainties in the 
parametrization of $B_m$ in Eq. (5), calculations are performed
with IQP matter for a rapidly increasing value of $B_m$ 
($\beta=0.02$ and $\gamma=3$) and for a slowly increasing value of $B_m$ 
($\beta=0.005$ and $\gamma=3$) taking $B_0=5\times 10^{18}$G. The 
respective maximum masses are $1.464M_{\odot}$ and $1.664M_{\odot}$. 
Apart from the uncertainties stemming from the parameters of the quark 
phase ($B$ and $\alpha_c$), the inadequate knowledge 
of the hadronic EOS at high densities relevant to a neutron star has important 
bearing on the phase boundaries in quark-hadron phase transition \cite{Mul}. 
To explore this effect, we have also performed calculations with 
a stiffer hadronic EOS \cite{Ser}. The phase boundaries for this
EOS (see Table I) are found to be shifted at much lower densities; 
the corresponding maximum masses are much larger than the observed
values. Recently similar conclusions have been drawn \cite{Mul} about 
the phase transition densities in an effective field theoretic model 
including non linear scalar and vector meson interactions which soften 
the hadronic EOS.

Neutron star cooling by URCA process may provide important informations 
about the interior constitution of the star. It is therefore of interest 
to see whether cooling by direct URCA process involving quarks can occur 
for star with quark matter. The decay of $d$ and $s$ quarks are 
kinematically allowed \cite{Iwa} if they satisfy the respective inequality 
conditions $p_F^u - p_F^e \leq p_F^d \leq p_F^u + p_F^e$ and
$p_F^u - p_F^e \leq p_F^s \leq p_F^u + p_F^e$. 
With $B_m=0$, direct URCA process involving $u$ and $d$ quarks 
is found to occur for $n_b \stackrel{>}{\sim} n_0$, while the latter relation
is satisfied for $n_b \stackrel{>}{\sim} 5.3n_0$. On the other hand, for 
$B_0 = 5\times 10^{18}$G ($\beta =0.01$ and $\gamma = 3$) the first
inequality condition is satisfied for $n_b \stackrel{>}{\sim} n_0$, while
the latter occurs for $n_b \stackrel{>}{\sim} 2.4n_0$. Thus
the threshold density for $s$ quark decay depends sensitively on $B_m$.
Large neutrino luminosity and therefore rapid cooling may serve as an
observational means for the presence of strong magnetic field in the 
inner core of the star, and work in this direction is in progress.


 \end{multicols}

\begin{references}
\bibitem{Ins} Even with this high magnetic field of $B_0 = 10^{19}$G, 
 the EOS is still found to be stable.
\end{references}
\end{document}